\documentclass[aip,reprint,article,amsmath,showpacs,rsi]{revtex4-1}

\usepackage{graphicx}
\usepackage{color}
\usepackage{hyperref}
\usepackage{soul}

\begin{document}

\newcommand{\MVAL}{190}

\title{High sensitivity photonic time-stretch electro-optic sampling of
  terahertz pulses}
\newcommand{\affiliationphlam}{Laboratoire PhLAM, UMR CNRS 8523, Universit\'e Lille 1,  Sciences et Technologies, 59655 Villeneuve d'Ascq, France}
\newcommand{\affiliationcerla}{Centre d'\'Etude Recherches et Applications (CERLA), 59655 Villeneuve d'Ascq, France}
\newcommand{\affiliationsoleil}{Synchrotron SOLEIL, L'Orme des Merisiers, Saint-Aubin, BP 48, 91192 Gif-sur-Yvette Cedex, France}
\author{C. Szwaj}
\affiliation{\affiliationphlam}
\affiliation{\affiliationcerla}
\author{C. Evain}
\affiliation{\affiliationphlam}
\affiliation{\affiliationcerla}
\author{M. Le Parquier}
\affiliation{\affiliationcerla}
\author{P. Roy}
\affiliation{\affiliationsoleil}
\author{L. Manceron}
\affiliation{\affiliationsoleil}
\author{J.-B. Brubach}
\affiliation{\affiliationsoleil}
\author{M.-A. Tordeux}
\affiliation{\affiliationsoleil}
\author{S. Bielawski}
\email{serge.bielawski@univ-lille1.fr} 
\affiliation{\affiliationphlam}
\affiliation{\affiliationcerla}

\begin{abstract} Single-shot recording of terahertz electric signals has
  recently become possible at high repetition rates, by using the {\it
    photonic time-stretch} electro-optic sampling (EOS) technique. However the
  moderate sensitivity of time-stretch EOS is still a strong limit for a range
  of applications. Here we present a variant enabling to increase the
  sensitivity of photonic time-stretch for free-propagating THz signals. A key
  point is to integrate the idea presented in Ref. [Ahmed et al.,
  Rev. Sci. Instrum. 85, 013114 (2014)], for upgrading classical time-stretch
  systems. The method is tested using the high repetition rate terahertz
  coherent synchrotron radiation source (CSR) of the SOLEIL synchrotron
  radiation facility. {The signal-to-noise ratio of our
    {terahertz digitizer could thus be straightforwardly
      improved by a factor $\approx 6.5$, leading to a noise-equivalent input
      electric field below $1.25$~V/cm inside the electro-optic crystal,
      over the 0-300~GHz band (i.e, 2.3~$\mu$V/cm/$\sqrt{\text{Hz}}$).} The sensitivity
    is scalable with respect to the available laser power, potentially
    enabling further sensitivity improvements when needed.}
 \end{abstract} 
\date{\today}
\pacs{41.60.Ap,29.27.Bd,05.45.-a}
\maketitle

\section{Introduction}
The so-called photonic time-stretch
technique~\cite{time_stretch_first_Coppinger_1999} enables single-shot
recording of electric pulses with picosecond
resolution~\cite{jalali_nature_photonics_review,time_stretch_transfer_function_han_2003,wong2011_time_stretch_balanced,han_2005_phase_diversity},
and high acquisition rates (tens of Mega pulses/seconds or more). The
speed of these systems has open the way to ``terahertz oscilloscopes''
with potential applications including analysis of electric signals,
guided millimeter waves~\cite{chang2000time}, freely propagating
terahertz pulses~\cite{roussel2015.EOS,roussel_these}, and
relativistic electron
bunches~\cite{roussel2015.EOS,roussel_these,roussel_anka_ibic2015}. However
speed is not the only important parameter of photonic time-stetch
digitizers. As for classical oscilloscopes, the sensitivity (i.e., the
minimum signal that can be distinguished from noise) also
largely determines the range of potential applications that may be
expected from photonic time stretch, and specific works have naturally
aimed at improving signal-to-noise
ratios~\cite{solli2008amplified,wong2011_time_stretch_balanced}.

{Besides, photonic time-stretch devices share common points with
  electro-optic sampling systems, which are used in terahertz
  physics~\cite{valdmanis1986subpicosecond,EOS_first_wu_1995,nahata1996wideband,EOS_first_spectral_encoding,sun1998analysis,peng2008optimal}
  and accelerator
  physics~\cite{PhysRevLett.88.124801,schmidhammer2009single,uvsor_eos_2012,PhysRevSTAB.15.070701,ANKA.EOS.IPAC13}. Hence
  photonic time stretch and electro-optic sampling can benefit from
  the advances from each other.

  Here we show that it is possible to design photonic systems that
  combine: (i) the ultra-high sentitivity which is available in
  recent state-of-art electro-optic
  sampling~\cite{ahmed2014detectivity,ahmed2014detectivity_pnas}, with
  (ii) the high repetition rate that is characteristic of photonic
  time-stretch~\cite{time_stretch_first_Coppinger_1999}. More
  specifically, we provide guidelines for increasing the sensitivity
  of photonic time-stretch setups employing a polarization modulation
  scheme~\cite{wong2011_time_stretch_balanced,roussel2015.EOS}. We
  test the efficiency of the strategy for the characterization of
  terahertz coherent synchrotron radiation (CSR) in the SOLEIL
  synchrotron radiation
  facility~\cite{evain2012.0295-5075-98-4-40006,roy.RSI.84.033102.2013,roussel2015.EOS,barros2015characteristics,tammaro2015high}.

\section{Experimental strategy}
\subsection{Brief review of related  photonic time-stretch and electro-optical sampling concepts} In order to record free-space terahertz
  wave electric fields with high repetition rates and high
  sentitivity, our experimental strategy is based on combining 
  several existing concepts, borrowed from photonic
  time-stretch and terahertz electro-optic sampling:

\subsubsection{Photonic time-stretch} We use photonic
  time-stretch~\cite{time_stretch_first_Coppinger_1999} which is a
  technique allowing electric field transients to be recorded with
  high repetition rates (the general principle is resumed in
  Fig.~\ref{fig:bib}a). Chirped laser pulses are modulated by the
  ultrafast electric pulses under investigation. Then the pulses are
  stretched in a long fiber until their duration is of the order of
  nanoseconds, and may be recorded by an oscilloscope. If some
  conditions are met~\cite{time_stretch_transfer_function_han_2003},
  the output pulses are replica of the input THz pulses, except they
  are slowed down by a factor:
\begin{equation}
M=1+\frac{L_2}{L_1},
\label{eq:stretch-factor}
\end{equation}
with $L_1$ and $L_2$ the lengths of the two fibers.

\subsubsection{Balanced detection} Electro-optic sampling with {\it balanced
  detection} is a well known noise-reduction technique, which is widely used
in terahertz time-domain
spectroscopy~\cite{valdmanis1986subpicosecond,nahata1996wideband,gaal2007measuring}. Association
of photonic time-stetch with balanced detection has been thus naturally
implemented in several
stups~\cite{wong2011_time_stretch_balanced,roussel2015.EOS}, and a conceptual
illustration is presented in Figure~\ref{fig:bib}b. In practice, a
quarter-wave plate is placed before the polarizing beam-splitter, and equal
powers are sent to the two balanced detector inputs.  This technique is
particularly efficient for decreasing the common mode noise (as laser relative
intensity noise, and amplified spontaneous emission noise of amplifiers placed
before the Pockels crystal).

    \subsubsection{Sensitivity increase using near-zero transmission point}
    In order to increase the sensitivity, a well-known approach consists in
    operating the set of polarizing elements {\it near zero transmission
      point}, e.g., placing the electro-optic crystal between near-crossed
    polarizers~\cite{jiang1999nearzeroopticaltransmissionpoint,shi2008theoretical}. An
    association of this technique with photonic time-stretch is presented in
    Figure~\ref{fig:bib}c. The responsivity (i.e., the laser power modulation
    amplitude per V/cm electric field) can be increased by orders of
    magnitude, if a large power is available. This strategy is widely used in
    single-shot diagnostics of electron bunches in electron accelerators and
    Free-Electron
    Lasers~\cite{PhysRevLett.88.124801,steffen2009electro,PhysRevSTAB.15.070701,ANKA.EOS.IPAC13}. Note
    however that associating this technique with balanced detection is not
    straightforward.

    \subsubsection{Balanced detection and high sensitivity} A further
    advance has been made recently, when Ahmed, Savolainen and
    Hamm~\cite{ahmed2014detectivity,ahmed2014detectivity_pnas}
    proposed a compact scheme that combines the advantages of the two
    previous methods (i.e., {near zero transmission point AND balanced
      detection}). This consists in inserting a partial polarizer
    (Brewster plates in
    Refs.~\cite{ahmed2014detectivity_pnas,ahmed2014detectivity}) in a
    classical balanced detection setup. We show here that this
    technique can be easily associated to a photonic time stretch
    setup, as is illustrated in Fig.~\ref{fig:bib}d (a more compact
    setup will be presented in the next section). Theoretically, for a
    given power on the balanced photodetector, and a given (weak) THz
    electric field, the output signal of Figure~\ref{fig:bib}d is
    expected to be larger than Figure~\ref{fig:bib}b's signal by a
    factor $a=1/\sqrt{T}$~\cite{ahmed2014detectivity}, with $T$ the
    power transmission of the Brewster plates set for the
    s-polarization (assuming perfect optics).

\begin{figure*}[htbp]
\centering\includegraphics[width=14cm]{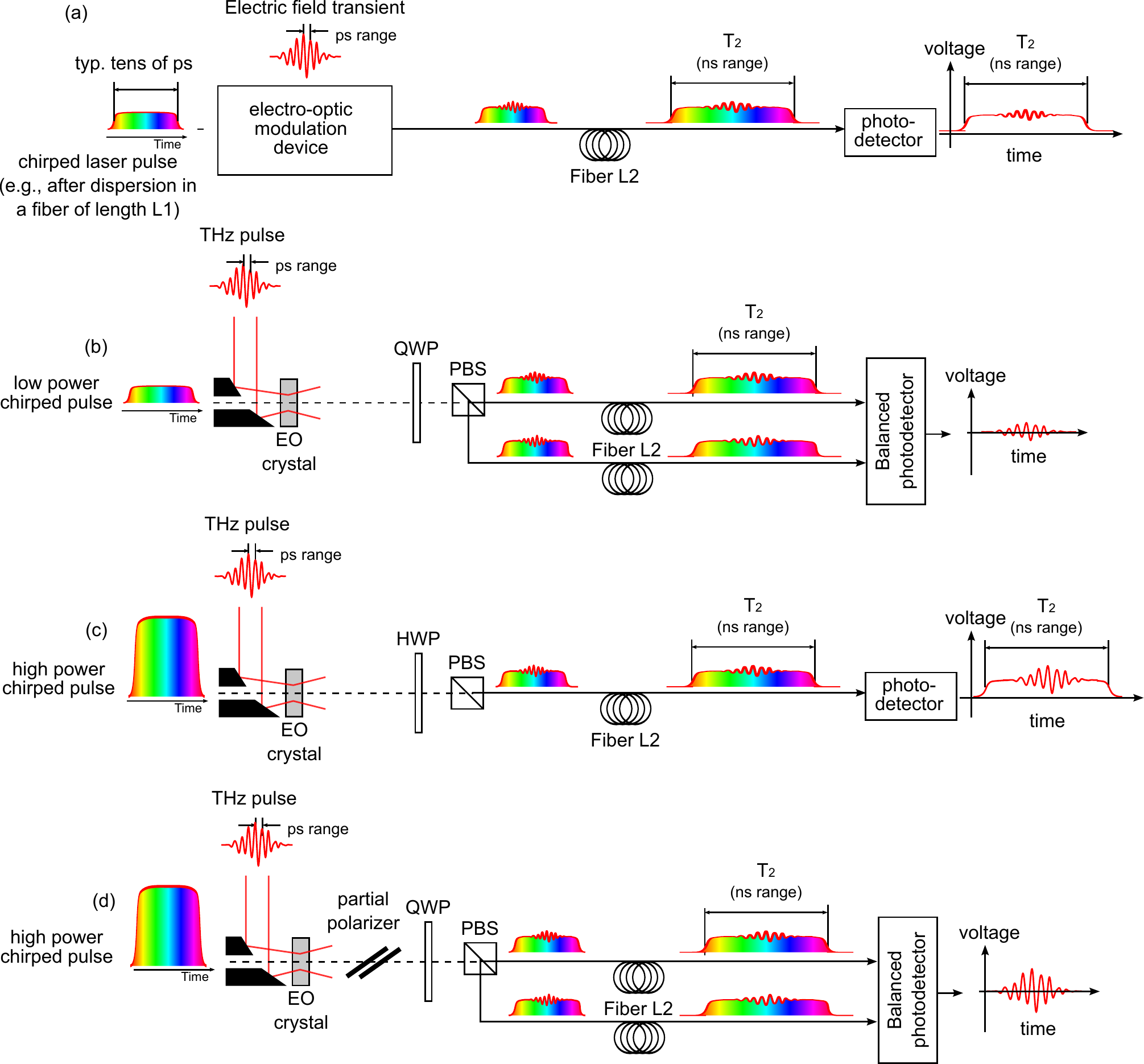}
\caption{{(a): Illustration of the photonic time
    stretch technique, and (b-d): possible associations with existing
    signal-to-noise enhancement techniques. (b): balanced detection,
    which reduces efficiently the common mode optical noise. (c): Near
    zero optical transmission point operation, which is particularly
    efficient for increasing the output electro-optic signal, if a
    sufficiently high power laser source is used. (d): Association of
    the three (a, b \& c) techniques: Balanced detection and near-zero
    transmission operation (using the trick from
    Ref.~\cite{ahmed2014detectivity}), and photonic time stretch. The
    detailed (and more compact) setup used in our experiments is
    displayed in Figure~\ref{fig:expsetup}.  PBS: polarizing
    beam-splitter. EO crystal: electro-optic crystal.}}
\label{fig:bib}
\end{figure*}

\subsection{Experimental setup}
We associated the photonic time stretch method and the Ahmed {\it et
  al.} high sensitivity strategy, in the compact system represented in
Figure~\ref{fig:expsetup}. Instead of using two fibers after a
polarizing beam splitter (as in Figure~\ref{fig:bib}d), we use a
single polarization-maintaining (PM) fiber for stretching the two
polarization states. The two components are separated only at the
fiber's output, using a fiber polarizing beam splitter (PBS in
Figure~\ref{fig:expsetup}). The stretch factor is $M=200$, i.e.,
6~GHz at the oscilloscope input corresponds to 1.2~THz at the
electro-optic crystal.

The orientations of the GaP crystal and of the laser and terahertz
polarizations are chosen according to
Ref.~\cite{PhysRevSTAB.15.070701}, in order to maximize the electric
field-induced birefringence (see Figure~\ref{fig:expsetup} for
details). Note that the electric field induced biferingence is at
45~degrees with respect to the [-1,1,0] axis (i.e., at 45 degres with
respect to the plane of the figure). The GaP crystal length is
$d=5$~mm. The laser is an Orange Ytterbium doped fiber laser from
Menlo Systems GMBH, delivering mode-locked pulses at 1030~nm, with
$\approx 30$~nm width, 20~mW average power, and 88~MHz repetition
rate. After selection by a pulse picker, the pulses are amplified
using a home-made Ytterbium Doped Fiber Amplifier (YDFA). The YDFA
output energy depends on the repetition rate, and is typically of the
order of $12$~nJ for the results presented hereafter. The laser
repetition rate is locked on the 104-th harmonic of the storage ring
revolution frequency (846.6~kHz) using an RRE synchronization system
from Menlo Systems.

Balanced detection is then performed between the two PBS outputs, using a
DSC-R412 amplified balanced detector (photoreceiver) from Discovery
Semiconductors, with 20~GHz bandwidth and 2800~V/W gain (specified at
1550~nm). An adjustable delay line, and a variable optical attenuator (VOA)
are used for matching the relative delays and powers of the two ports,
respectively (see Appendix~\ref{sec:appendix_balanced} for delay adjustment
procedure and noise suppression by balanced detection). The signals are
acquired using a Lecroy Labmaster 10~Zi oscilloscope, with an
(overdimensionned) 36~GHz bandwidth. The data are low-pass filtered to 6~GHz,
which corresponds to a bandpass limitation of 1.2~THz at the crystal
location (see Appendix for details on the transfer function).

In addition to the laser pulses interacting with the THz signals,
extra laser pulses are also generated for providing a dark reference
(i.e., without terahertz signal), which is subtracted from the data,
at the processing stage.

Tests of the EOS setup are performed using coherent THz pulses that
are delivered by the Synchrotron SOLEIL facility, in normal user mode,
at the AILES beamline~\cite{roy.IPT.49.139.2006}. The conditions are
similar to the ones of Ref.~\cite{roussel2015.EOS}, except that the
beam current is lower here (12~mA), and 8 electron bunches are
operated simultaneously.  Below, we focus on the emission by one of
the 8 electron bunches, at a repetition rate of 0.85~MHz.

\begin{figure}
\centering\includegraphics[width=8.5cm]{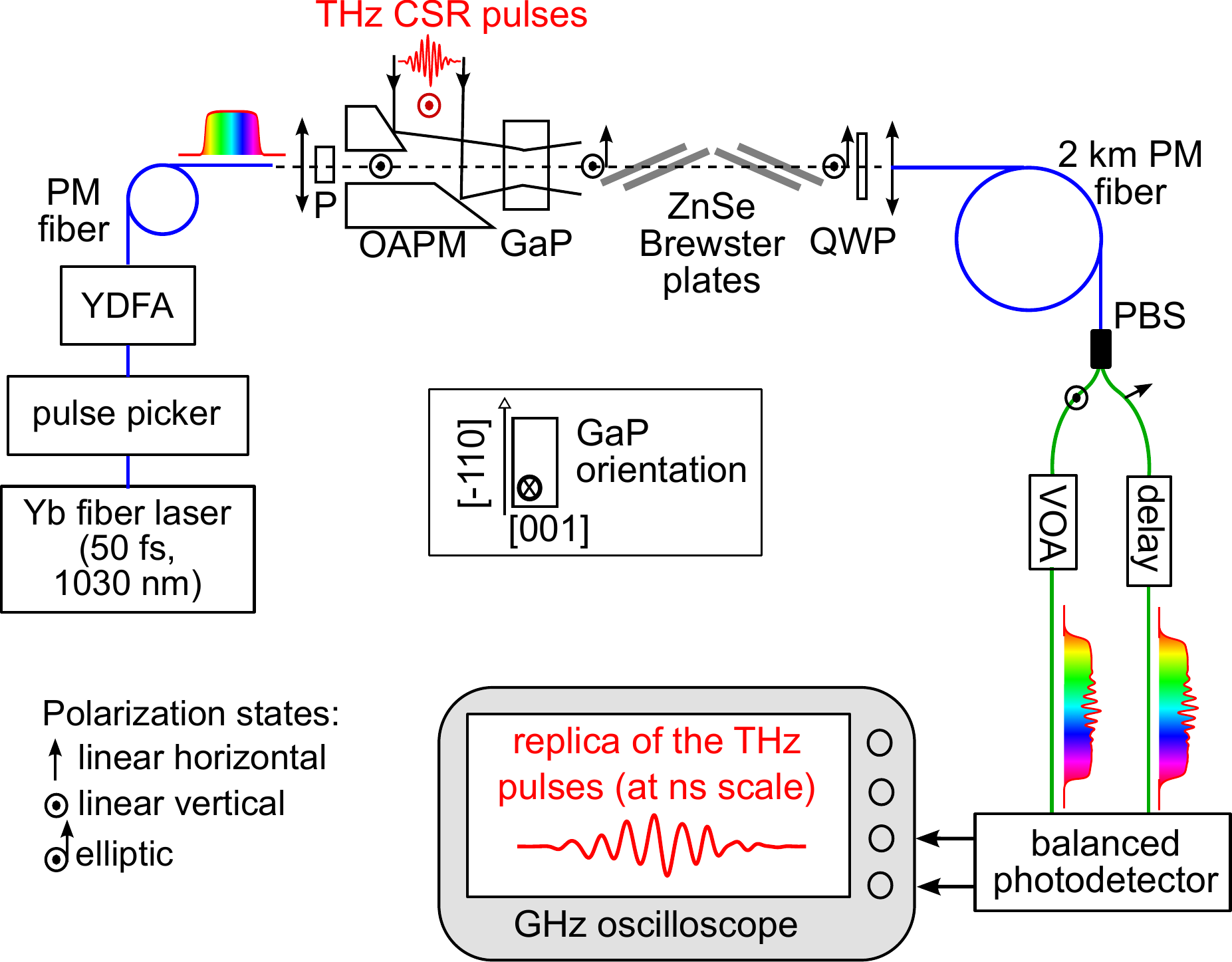}
\caption{Experimental setup, combining high sensitivity electro-optic
  sampling and photonic time-stretch. High sensitivity is obtained by
  using the four ZnSe plates, using the arrangement proposed in
  Ref.~\cite{ahmed2014detectivity}. Blue and green lines represent
  polarization maintaining (PM), and non-polarization maintaining (SM)
  fibers, respectively. YDFA: Ytterbium-doped fiber amplifer
  {delivering 12~nJ pulses,} P: polarizer, OAPM: 50.8 focal length
  off-axis parabolic mirror. GaP: Gallium Phosphide electro-optic
  crystal. QWP: quarter-wave plate (oriented at 45 degrees with
  respect to the figure plane), VOA: variable optical attenuator. PBS:
  fibered polarizing coupler (the slow axis of the PM fiber is coupled
  to the VOA port). The two fiber collimators have 11~mm focal
  length. The vertical and horizontal polarization components are
  injected in the slow and fast axes of the PM fiber, respectively. }
\label{fig:expsetup}
\end{figure}

\section{Experimental results}
In Figure~\ref{fig:typ_exp_signals}, we have represented a typical raw
single-shot signal, i.e., the signal from the balanced photodetector
over a 6~GHz bandwidth. From these data, we can deduce the evolution
of the THz electric field value inside the crystal (see
Appendix~\ref{sec:modeling} for details), and the evolution of the THz
field-induced birefringence phase shift
(Fig.~\ref{fig:typ_exp_signals_Vcm}).

The terahertz emission occurs in the form of bursts, which are
separated by $1.25$~ms, and contain several hundreds of THz
pulses. The upper parts of Figs.~\ref{fig:typ_exp_signals},
and~\ref{fig:typ_exp_signals_Vcm} display recordings used with the
present setup, and the lower part displays a reference EOS signal,
obtained without the ZnSe plates. Panels a-c represent the same data
in different ways: (a) are examples of single-shot signals, (b) and
(d) represent series of EOS signals $V_n(t)$ over many shots $n$ (at
846.6~kHz repetition rate), and (c) is a vertical cut of (b). In
both cases (with and without ZnSe plates), the laser power at the EOS
setup input has been set as high as possible: {(i) just below the detector
  saturation for the reference without ZnSe plates -- i.e., ensuring
  the best signal to noise ratio (SNR)-- and (ii) $\approx$4 times
  below detector saturation} in the setup with ZnSe plates (a value
limited by the available amplifier's output power), i.e., leading to a
not optimal value for the SNR.

A significant difference in SNR is visible between these two
recordings (compare in particular the (b) and (c) panels of the upper
and lower data). Remarkably, the SNR is better with ZnSe plates, even
though the laser power on the photodiodes is significantly lower. This
indicates that further improvement in SNR can even be obtained in the
future, just by increasing the output power of our amplifier (by a
factor 4).

\section{Discussion}
\subsection{Increase in signal level}
{For a given power at the detector's inputs, the signal is expected to increase
by a factor (see appendix and reference~\cite{ahmed2014detectivity}):
\begin{eqnarray}
a=1/\sqrt{T},
\end{eqnarray}
where $T$ is the transmission of the ZnSe plates for the
s-polarization. Assuming a refractive index $n=2.4858$ for ZnSe (at
1030~nm), and perfect Brewster incidence, the transmission of each
brewster interface is 48\% for the s-polarization, and the increase
factor is hence $a=18.9$.}

{In the present experiment, the signal enhancement on the photodetectors is
  lower than this value {because the output laser power is
    $\approx$4 times lower than without ZnSe plates}. Hence a factor $\approx 4.8$ in
  signal enhancement (instread of 18.9) is actually obtained. Note however
  that the increase in signal to noise ratio is higher than this value,
  because the shot-noise is also lowered.}

\begin{figure*}
\centering\includegraphics[width=12cm]{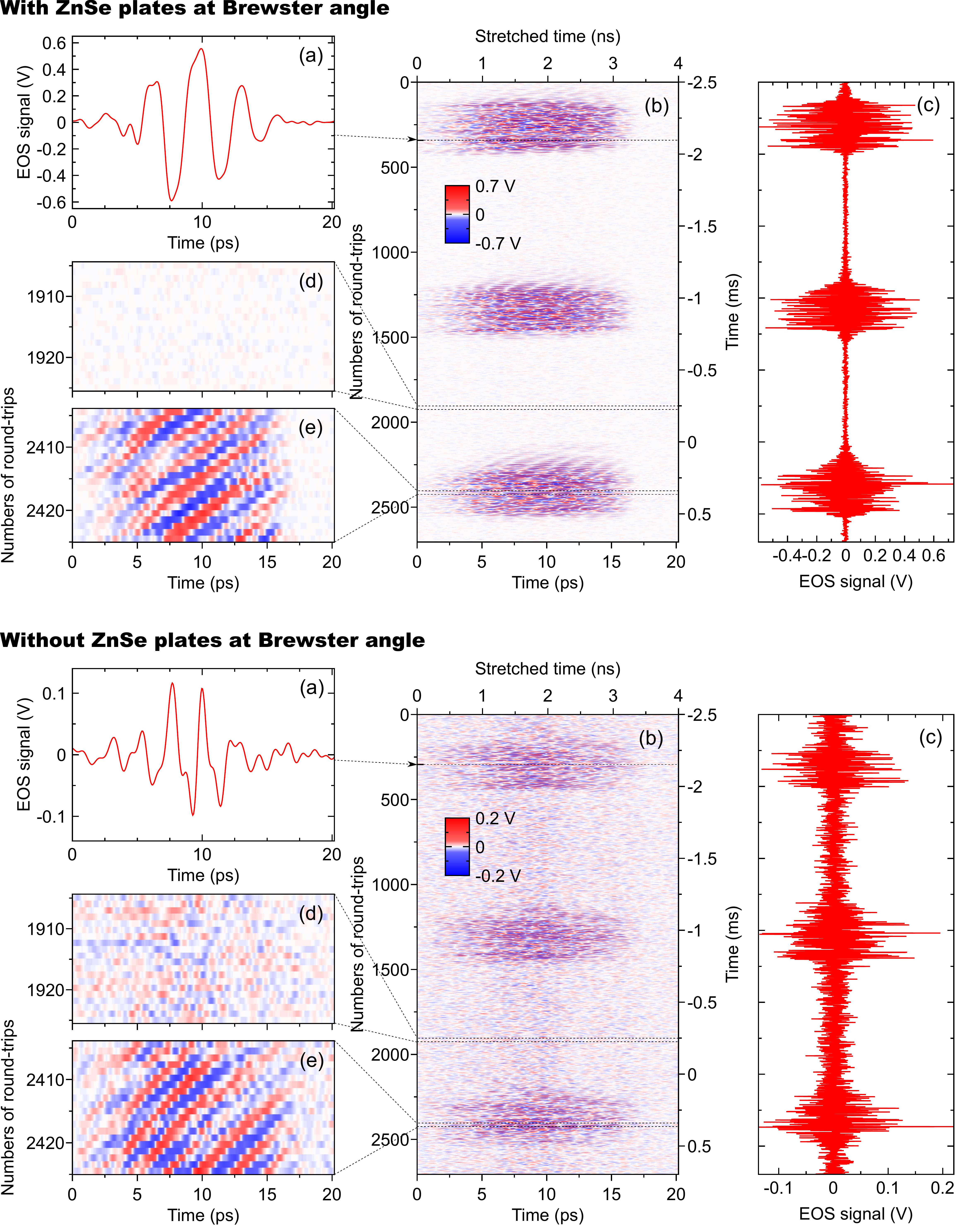}
\caption{Single-shot recordings (balanced photodetector voltage) of
  successive THz coherent pulses (CSR). The upper part corresponds to
  a recording using the present setup. The lower part coresponds to
  the setup without sensitivity enhancement system (ZnSe plates). (a):
  one single shot EOS signal, (b) and (d): series of EOS THz signals
  $V_n(t)$ over many round-trips $n$, represented as a colorscale
  diagram [(d) is a zoom of (b)]. (c): signal at center
  $V_n(t=10\;\textrm{ps})$ versus round-trip number $n$. Note the
  difference in SNR ratio between upper and lower data.}
\label{fig:typ_exp_signals}
\end{figure*}

\begin{figure*}
\centering\includegraphics[width=12cm]{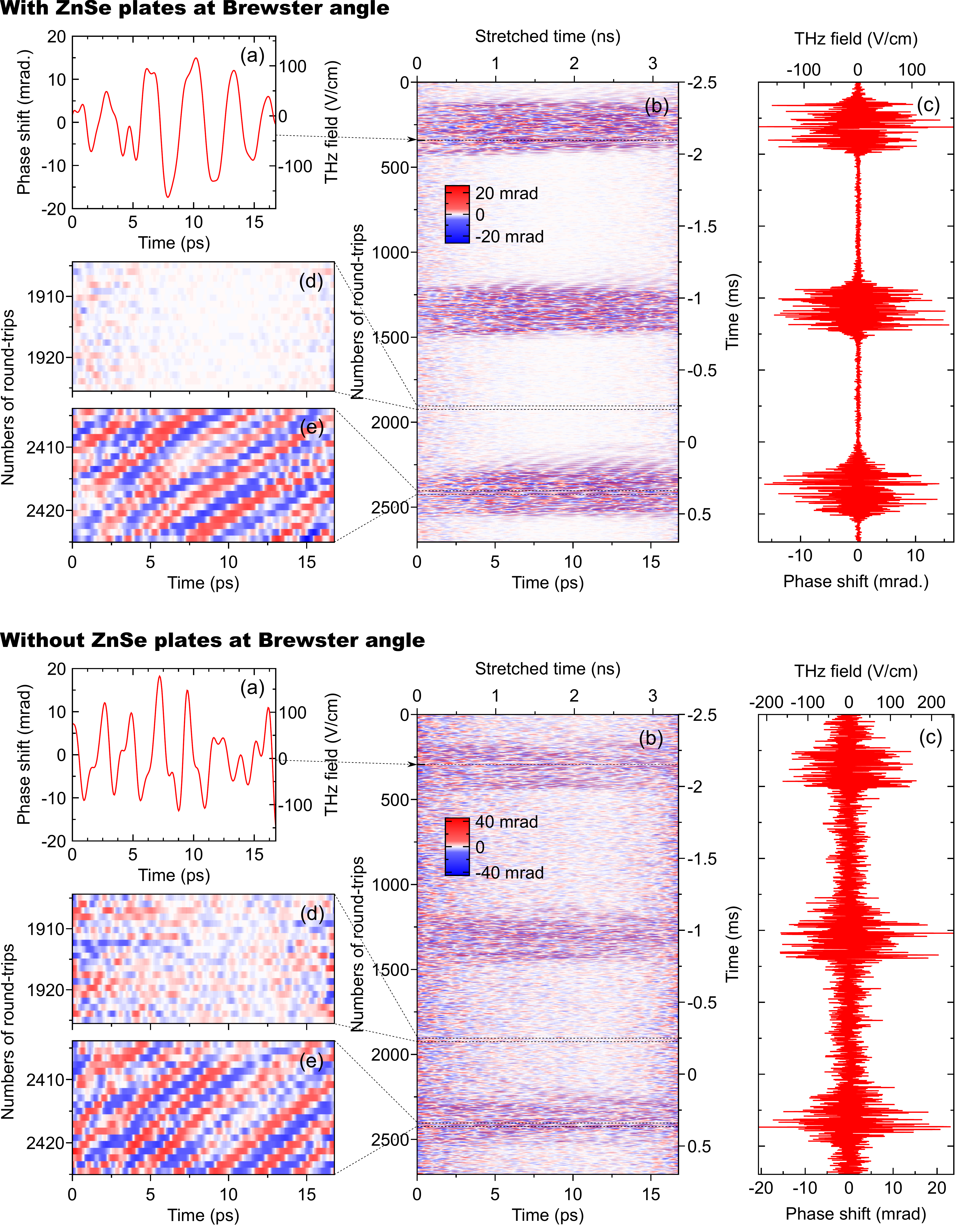}
\caption{{Estimation of the} phase-shift inside the
  crystal, and of the THz electric field from the data of
  Figure~\ref{fig:typ_exp_signals}. This estimation assumes a constant
  electro-optic coefficient $r_{41}=1$~pm/V\cite{nelson1968electro}
  for Gallium Phosphide over the THz signal
  bandwidth. {The 3~dB system bandwidth is 0.4~THz}
  (see Appendix~\ref{sec:modeling}).}
\label{fig:typ_exp_signals_Vcm}
\end{figure*}

\subsection{Enhancement of SNR and noise-equivalent input electric field} {The
  sensitivity of this type of detector can be characterized by the input THz
  signal, which would be equivalent to the measured noise. This
  noise-equivalent input signal is non-stationary (i.e., its statistical
  properties depends on time) because the laser power varies with time. This
  is also complicated by the transfer function of the photonic time-stretch
  system (see Figure~\ref{fig:transfer_function} and
  Appendix~\ref{sec:modeling}).}

{It is nevertheless possible to define a simple estimate for
  the noise level, that can be used for comparison with other setups (and
  leave detailed analysis of the non-stationary noise for future works).}  For
this, we use data without THz signal and calculate the RMS noise fluctuation
estimate over quasi-flat region of the transfer function, up to 300~GHz (see
Figure~\ref{fig:transfer_function}). We also assume a flat frequency response
for the $r_{41}$ coefficient on this bandwidth. The RMS noise estimated using
this definition} is displayed in Figure~\ref{fig:exp_noise_versus_time}.

{The minimum input noise value is of the order of
  1.25~V/cm  (RMS, in the crystal) over the 0-300~GHz band, i.e.,
  2.3~$\mu$V/cm/$\sqrt{\text{\text{Hz}}}$. This corresponds to an
  improvement by a factor $\approx$6.75 with respect to the balanced
  detection data otained without ZnSe plates.}

\begin{figure}
\centering
\includegraphics[width=8.5cm]{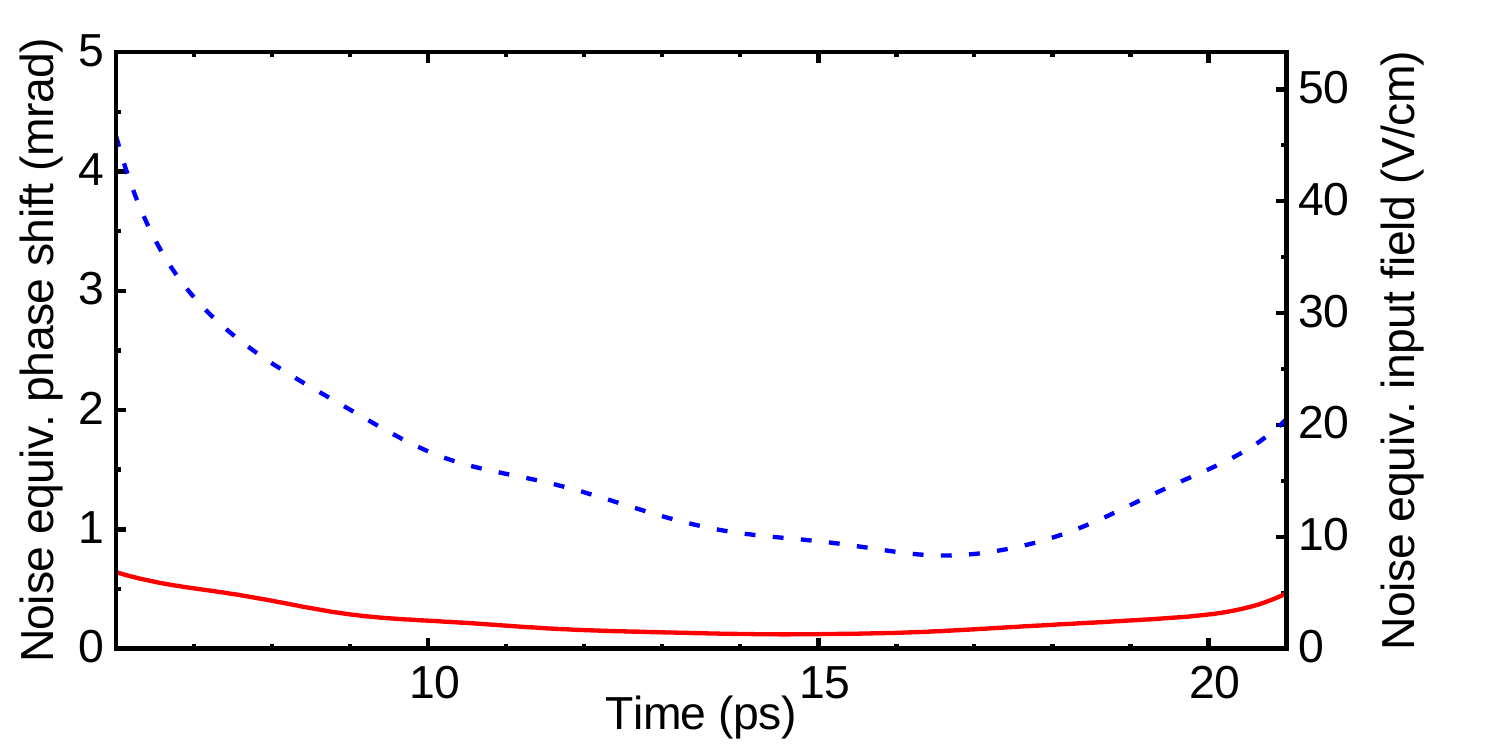}
\caption{Input electric field equivalent to the noise level, over a
  bandwidth of 300GHz. The red (full line) and blue (dashed line) lines
  correspond to the setup with and without the Brewster plate system,
  respectively. The left vertical axis corresponds to the THz-induced
  birefringence phase shift.}
\label{fig:exp_noise_versus_time}
\end{figure}


\begin{figure}
\centering\includegraphics[width=8cm]{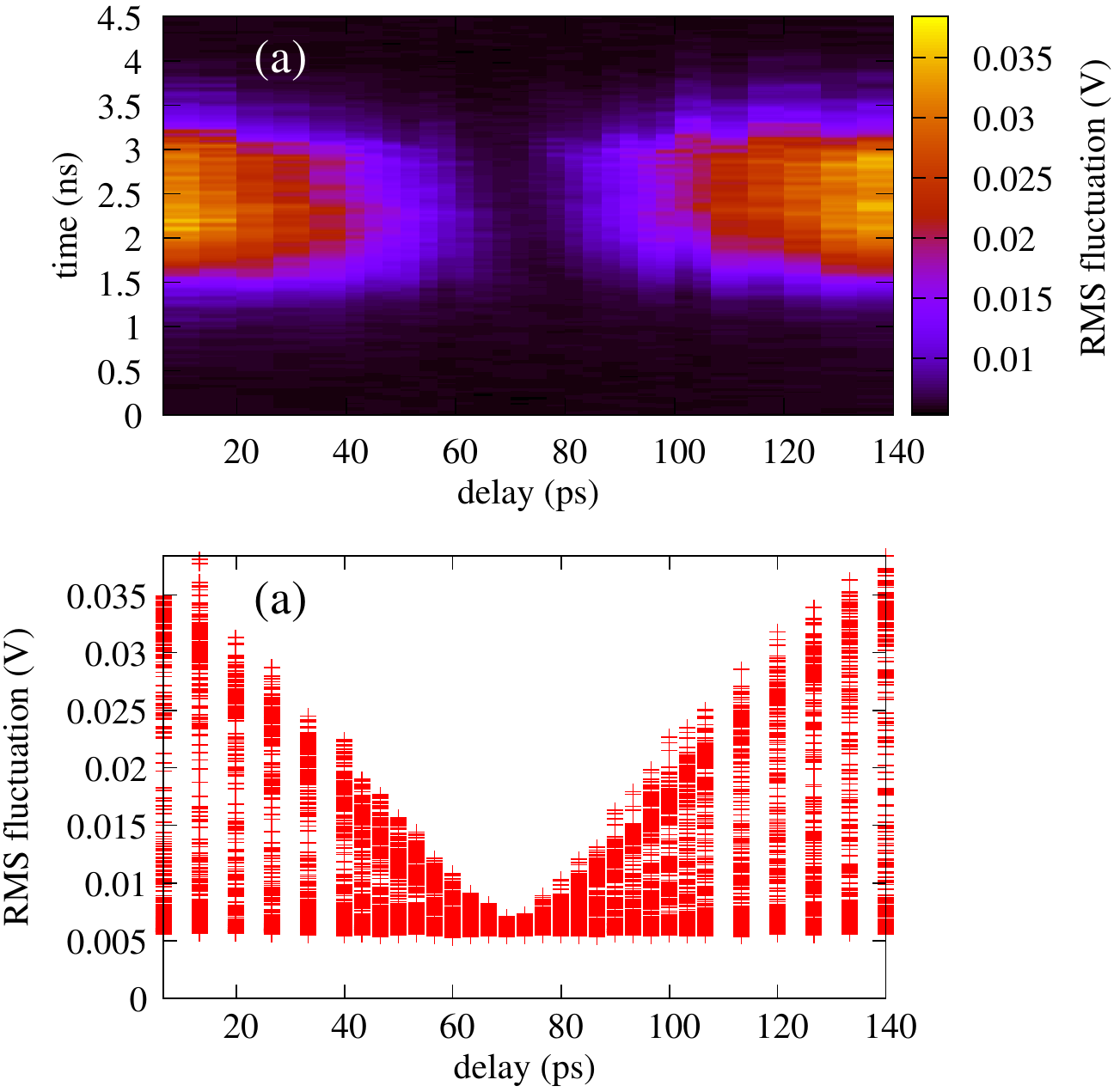}
\caption{Measured output noise level (RMS oscilloscope signal in absence of
  THz pulses), versus delay between the two balanced photodetector inputs
  (note that the origin of the horizontal axis is arbitrary). (a): noise
  versus time at oscilloscope input (vertical axis), and delay (horizontal
  axis). (b): same data projected vertically (ie., RMS noise levels for each
  oscilloscope times, versus delay). Laser noise cancellation is clearly
  visible when the delay line adjustment is at the $70$~ps value (i.e.,
  corresponding to zero delay between the two balanced detection
  signals). Detector noise (in absence of laser) corresponds to $\approx 5.8$~mV~RMS.
  Noise is measured on a 6~GHz bandwidth, corresponding to an 1.2~THz
  limitation in the electro-optic crystal.}
\label{fig:exp_noise_vs_delay}
\end{figure}

\section{Conclusion} 
Relatively simple photonic time stretch digitizers for free-propagating
electric fields can be designed with balanced detection and ``near
extinction'' operation, thus leading to a significant improvement of the
signal to noise ratio (by factor 6.7 here). Further increase of the amplifier
power output up to the optimal level for the photodiodes (i.e., by a factor 4)
should directly lead to the SNR enhancement close to the value $a=1/\sqrt{T}=
18.9$. In addition, as for scanning EOS~\cite{ahmed2014detectivity}, further
increase of the detectivity should be made possible by decreasing the
transmission $T$ of the Brewster set plates for the (s) polarization, and
increasing the laser power by the same factor.

This technique can be directly applied when the modulation is performed on the
birefringence of an electro-optic crystal. Hence -- in addition to the
detection of THz fields -- this also opens new ways for improving the SNR in
electron bunch shape measurements systems for accelerators and free-electron
lasers. Further relevant works may concern the association of the present
scheme with the so-called amplified time-stretch
strategy~\cite{solli2008amplified} (with the aim to further enhance
detectivity), {as well as extending the THz detection
  bandwidth}, using e.g., phase
  diversity~\cite{tarighat2007two,han_2005_phase_diversity,chang2000time}. Last
  but not least, reduction of the setup cost may be possible, e.g., by
  replacing the PM fiber with an SM fiber if a system with circulators can be
  implemented efficiently~\cite{fard2011spectral}. Applications of this method
  to digitization of cable-transported
  signals~\cite{time_stretch_first_Coppinger_1999,wong2011_time_stretch_balanced}
  still remains an open question, as the feasibility can depend on the
  possibilities to apply the Ahmed-Savolainen-Hamm principle to fibered
  modulators (e.g., polarization
  modulators~\cite{wong2011_time_stretch_balanced}).

\section{Acknowledgements}
We would like to thank Menlo Systems for important advices. The work has
been supported by the BQR of Lille University (2015). The work has
also been supported by the Ministry of Higher Education and Research,
Nord-Pas de Calais Regional Council and European Regional Development
Fund (ERDF) through the Contrat de Projets \'Etat-R\'egion (CPER)
20072013, and the LABEX CEMPI project (ANR-11-LABX-0007). Preparation
of the experiment used HPC resources from GENCI TGCC/IDRIS
(i2015057057, i2016057057).

\appendix
\section{Note on the efficiency of balanced detection}
\label{sec:appendix_balanced}
Proper adjustment of the delay line is expected to lead to an efficient noise
cancellation between the two balanced detector's inputs. This is illustrated
in Figure~\ref{fig:exp_noise_vs_delay} where we represented the noise level
(at the balanced detector's output) versus delay, in absence of THz
signal. The remaining noise is slightly higher than the ``no-signal'' detector
noise ($\approx 5.8$~mV RMS over the 6~GHz bandwidth) . This indicates a
satisfying common mode rejection on the whole temporal window defined by the
laser pulse duration.
\section{Modeling details} 
\label{sec:modeling}

\subsection{Output signal versus input THz electric field} 
The experimental setup contains two main building blocks, which have
been theoretically modeled in previous works. In this appendix, we
provide for this particular setup: (i) the calculation os the output
signal for a given phase modulation in the crystal, and (ii) the
bandwidth limitation due to the time-stretch (or spectral encoding)
strategy.

The EOS principle (Fig.~\ref{fig:expsetup}) can be divided in two parts: (i)
The Pockels effect, that leads to the modulation of the two polarization
components injected in the PM fiber (the components are eventually separated
by the polarizing beam-splitter), and (ii) the dispersion in the long PM
fiber. Based on the corresponding litterature on electro-optic sampling and
photonic time stretch~\cite{time_stretch_first_Coppinger_1999,time_stretch_transfer_function_han_2003,wong2011_time_stretch_balanced,han_2005_phase_diversity},
each part can be modeled in a relatively straightforward way.

The THz electric field $E_{THz}(t)$ induces a birefringence, whose optical
axes are oriented at 45 degrees with respect to the laser polarization. In the
horizontal-vertical axes frame, the Jones matrix of the electro-optic
crystal is~\cite{casalbuoni2008numerical}
\begin{equation}
{\bf M}_{GaP}=\exp(i\phi_0)\left( \begin{array}{cc}
    \cos{\frac{\Delta\phi}{2}}&i\sin{\frac{\Delta\phi}{2}} \\
    i\sin{\frac{\Delta\phi}{2}} & \cos{\frac{\Delta\phi}{2}}
\end{array} \right),
\end{equation}
where $\Delta\phi(t)$ represents the relative phase retardation
between the two optical axes of the crystal, and $\phi_0$ a constant
phase shift. {Note that, at a given THz frequency~\cite{casalbuoni2008numerical}:
\begin{equation}
\Delta\phi(t)=\frac{2\pi d}{\lambda}n_0^3r_{41}E_{THz}(t),
\end{equation}
for the present crystal orientation, with $E_{THz}(t)$ the THz field
inside the crystal, $r_{41}$ the relevant electro-optic coefficient at the
considered THz frequency,} $\lambda$ the laser wavelength in vacuum,
$n_0$ the crystal refractive index, and $d$ its thickness. Here:
$\lambda=1030$~nm, $n_0=3.11$\cite{parsons1971far}, and $r_{41}\approx
1$~pm/V~\cite{nelson1968electro}.

At the input of the PM fiber, the two laser polarization components
$E_1$ and $E_2$ are: (see, e.g., Ref.~\cite{ahmed2014detectivity} for details):

\begin{equation}
 \left( \begin{array}{c}
E^{in}_x(t) \\
E_y^{in}(t)   
\end{array} \right)
=
\frac{\sqrt{T}E_0(t)}{\sqrt{2}}\left( \begin{array}{c}
\cos{\frac{\Delta\phi(t)}{2}}-\frac{1}{\sqrt{T}}\sin{\frac{\Delta\phi(t)}{2}}\\
i\left[\cos{\frac{\Delta\phi(t)}{2}}+\frac{1}{\sqrt{T}}\sin{\frac{\Delta\phi(t)}{2}}\right]
\end{array} \right),\label{eq:EOmodulation}
\end{equation}
where we have omitted the global constant phase shift due to
propagation, $T$ is the transmittance of the Brewster plate set for
the s-polarization (perfect transmission is assumed for the other
polarization). $E_0(t)$ represents the complex envelope of the chirped
laser pulse just before the crystal.

Then each polarization component experiences dispersion in the PM
fiber. {Neglecting high-order dispersion, losses, and
  nonlinearity}, the two components $E_{x}(z,t)$ and $E_{y}(z,t)$
satisfy the following propagation equation:
\begin{eqnarray}
&\frac{\partial E_{x}}{\partial z}=-i\frac{\beta_{2x}}{2}\frac{\partial^2 E_{x}}{\partial t^2}\label{eq:dispersionx}\\
&\frac{\partial E_{y}}{\partial z}=-i\frac{\beta_{2x}}{2}\frac{\partial^2 E_{x}}{\partial t^2},\label{eq:dispersiony}
\end{eqnarray}
where $\beta_{2x}$ and $\beta_{2y}$ are the dispersions for each
polarization components of the PM fiber. However, the difference of
dispersion has negligible effect in the present experiment. Hence we
consider that each component propagates with the same dispersion
$\beta_{2x}=\beta_{2y}=\beta_2$ and a same equivalent length $L_x=L_y=L$.

Simulation of the EOS system hence requires the integration of
Eqns.~(\ref{eq:dispersionx},\ref{eq:dispersiony}), where the initial
condition (at $z=0$) is given by Eq.~(\ref{eq:EOmodulation}). Then, we easily
deduce the powers $| E_{x}|^2$ and $| E_{y}|^2$ on the two balanced photodiodes.

\subsection{Bandwidth limitations} 
\begin{figure}[htbp]
\centering\includegraphics[width=8cm]{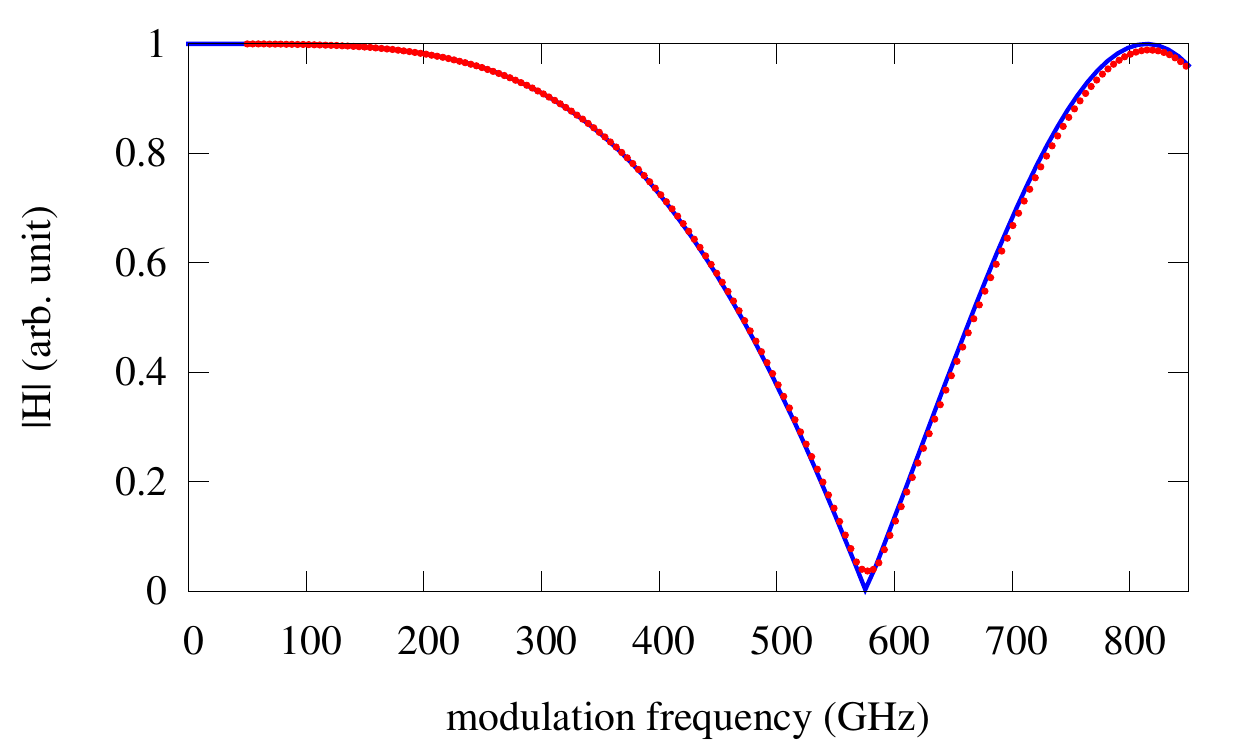}
\caption{Transfer function (more precisely $|H(f_m)|$). Red dots:
  numerical result using integration of
  Eqns.~(\ref{eq:dispersionx},\ref{eq:dispersiony}), with
  $\Delta\phi(t)=a_m\cos{2\pi f_mt}$ . Blue curve: Analytic result
  using Eq.~(\ref{eq:transfer_function_analytic}).
  $\beta_2=24$~ps$^2$/km, $L_1=10$~m, $L2=2000$~m, $a_m=0.01$. The
  initial laser pulse is Gaussian, with 40~fs duration (FWHM).}
\label{fig:transfer_function}
\end{figure}

{The frequency response can be stutied by computing the complex
amplitude of the balanced signal, when the input signal is sinusoidal:
\begin{eqnarray}
\Delta\phi(t)=a_m\cos 2\pi f_mt.
\end{eqnarray}
This can be performed by numerically after integrating
Eqns.~(\ref{eq:dispersionx},\ref{eq:dispersiony}), using
Eq.~(\ref{eq:EOmodulation}) as the initial condition. Furthermore,
analytical expressions for the transfer function may also be
obtained. The situation described by Eq.~\ref{eq:EOmodulation} is
formally analog to the modulation by a Mach-Zehnder modulation with
zero chirp~\cite{time_stretch_transfer_function_han_2003}, and thus we
expect the transfer function to be identical:}
\begin{eqnarray}
H(f_m)&&=\cos\left(2\pi^2\beta_2\frac{L_2}{M}f_m^2\right)\\
&&\approx \cos\left(2\pi^2\beta_2L_1f_m^2\right),\label{eq:transfer_function_analytic}
\end{eqnarray}
assuming $a_m$ small and $M$ large. $H(f_m)$ is the complex amplitude of the balanced detection signal,
normalized to its value at $f_m=0$ (and where a global constant phase
shift has been omitted). The transfer function is represented in
Figure~\ref{fig:transfer_function} with the parameters of the
experiment. $|H(f_m)|$ presents a fading phenomenon, which is
characteristic of photonic time-stretch, and whose cutoff frequency
depends on the length of the fiber ($L_1$). The first zero is located
at:
\begin{eqnarray}
f_{m0}\approx\frac{1}{\sqrt{4\pi\beta_2L_1}}.
\end{eqnarray}
In time domain, this fading corresponds the well-known trade-off between the resolution
$\tau_R$ and time window $\tau_w$ of the analysis~(see e.g., Ref.~\cite{sun1998analysis}, as well as Ref.~\cite{peng2008optimal} for a detailed analysis):
\begin{eqnarray}
  \tau_R=O(\sqrt{\tau_w\times\tau_{L0}}),
\end{eqnarray}
where $\tau_{L0}$ is the unstretched laser pulse duration, and $\tau_w\approx
\beta_2L_1$ if $\tau_w\gg\tau_{L0}$.

\subsection{Computation of the THz electric field from experimental
  data}
{In the frequency region where the transfer function
  $H$ is close to 1 (i.e., over the detector bandpass), and assuming
  that $r_{41}$ is almost constant, we can easily deduce the terahertz
  electric field evolution $E_{THzn}(t)$, and the birefringence
  phase-shift inside the crystal $\Delta\phi_n(t)$ at electron bunch passage  $n$ from
  the experimental data. For small values of $\Delta\phi_n(t)$:
\begin{eqnarray}
  \Delta\phi_n(t)\approx\frac{1}{a}\frac{V_{1n}(t)-V_{2n}(t)}{V_1^0(t)+V_2^0(t)}\\
  E_{THzn}(t)=\frac{\lambda}{2\pi d n_0^3r_{41}}\Delta\phi_n(t),
\end{eqnarray}
where $V_{1n}(t)-V_{2n}(t)$ is the raw balanced detection signal
(corresponding to Figure~\ref{fig:typ_exp_signals}), $V_1^0$ and
$V_2^0$ are the individual signals corresponding to each detector
without THz field ($V_1^0$ and $V_2^0$ are identical). $a$ is the
signal enhancement factor defined above (and $a=1$ without ZnSe
plates). Data presented in Figures~\ref{fig:typ_exp_signals_Vcm}
and~\ref{fig:exp_noise_versus_time} are obtained using these
assumptions (in particular, we did not attempt to deconvolve the data
using informations on the transfer function $H$).}


\bibliography{high_sensitivity_ts_eos1}
%
\end{document}